\documentclass[aps,prl,reprint,groupedaddress]{revtex4-1}
\usepackage{graphicx}

\begin{document}

\title{BaCu$_3$O$_4$: High Temperature Magnetic Order in One-Dimensional S=1/2
Diamond-Chains}

\author{C. W. Rischau$^1$, B. Leridon$^1$$*$, D. Colson$^2$, A. Forget$^2$ and P. Monod$^1$}
\affiliation{
$^1$Laboratoire de Physique et d'Etude des Mat\'eriaux,  UMR8213- CNRS-ESPCI ParisTech-UPMC \\ 10 rue Vauquelin, 75231 Paris Cedex 05, France\\
$^2$Service de Physique de l'Etat Condens\'e, Orme des Merisiers, IRAMIS, CEA-Saclay
(CNRS URA 2464), 91191 Gif sur Yvette Cedex, France\\
$*$ Corresponding author: brigitte.leridon@espci.fr}

\date{\today}

\begin{abstract}
The magnetic properties of the alkaline earth oxocuprate BaCu$_3$O$_4$ are investigated. We show that the characteristic Cu$_3$O$_4$ layers of this material can be described with diamond chains of antiferromagnetically coupled Cu 1/2 spins with only a weak coupling between two adjacent chains. These Cu$_3$O$_4$ layers seem to represent a so far unique system of weakly coupled one-dimensional magnetic objects where the local AF ordering of the Cu$^{2+}$ ions leads to an actual net magnetic moment of an isolated diamond chain. We demonstrate a magnetic transition at a high N\'eel temperature $T_{N}=336$ K.
\end{abstract}

\maketitle

\section{\label{intro}Introduction}

Lamellar copper oxides have been extensively studied not only regarding high-critical-temperature ($T_c$) superconductivity, but also regarding low-dimensional quantum magnets. Cupric oxide (CuO) itself exhibits two antiferromagnetic (AF) transitions at N\'eel temperatures $T_{N1,N2}=213, 230$ K denoting a strong superexchange energy (70 meV)  of the Cu-O-Cu coupling  \cite{Yang:1989}. Recently, Kimura \textit{et al.} detected ferroelectricity in CuO  for $T_{N1}<T<T_{N2}$  and thus identified CuO as an induced-multiferroic \cite{Kimura:2008}. Cu-O is also the key component of high-$T_c$ cuprate superconductors like YBa$_2$Cu$_3$O$_{x}$ (YBCO), (La,Sr)$_2$CuO$_4$ or HgBa$_2$Cu$O_{4+\delta}$, which are composed of CuO$_2$ layers.  These materials exhibit an antiferromagnetic phase in the underdoped non-superconducting regime of the phase diagram, but some groups additionally reported the existence of a commensurate alternate magnetization - probably of orbital nature - in the Cu-O planes of YBCO and HgBa$_2$Cu$O_{4+\delta}$ well in the doping range of superconductivity \cite{Fauque:2006,Mook:2008,Li:2008,Li:2010}. Another interesting family of lamellar copper oxides are the oxychlorures like Ba$_2$Cl$_2$Cu$_3$O$_4$ or Sr$_2$Cl$_2$Cu$_3$O$_4$ (2234 materials) which are composed of  Cu$_3$O$_4$ planes interseeded with Ba and Cl layers. The magnetic structure of these Cu$_3$O$_4$ layers results in two AF ordering temperatures and has been the focus of extensive experimental and theoretical studies (see \cite{Ito:1997,Chou:1997} and Refs therein).\\
Here we report on the magnetic properties of the alkaline earth oxocuprate BaCu$_3$O$_4$ which is composed of Cu$_3$O$_4$ layers \cite{Bertinotti:1989,Yakhou:1996} with a different structure than those of the 2234 materials, and, as will be demonstrated, very different magnetic properties.  As a matter of fact, BaCu$_3$O$_4$ exists only as a parasitic phase in YBCO samples \cite{Bertinotti:1989}, or on thin films grown on top of a suitable, epitaxial perovskite buffer layer \cite{Samoylenkov:1999}. We therefore performed our measurements on YBCO polycrystalline samples and single crystals which were shown to contain BaCu$_3$O$_4$.

\section{\label{exp}Methods}
The magnetization of thirteen polycrystalline and six single crystalline YBCO samples with various oxygen contents (see Tab. \ref{table1}) was investigated using a MPMS 5T SQUID magnetometer. The polycrystalline samples were prepared at the SPEC-CEA from three different batches of YBa$_2$Cu$_3$O$_{x}$ powder (batches A, B and C) by compacting the powder into cylinders of 6 mm diameter and 6 mm height and annealing them under appropriate N$_2$-O$_2$ mixtures in order to obtain the desired oxygen content. Four different disks of YBCO single crystals (batches D, E, F, G) with various oxygen contents were obtained from different sources and cut into cubes with an edge length of usually 5 mm. The oxygen content of the samples D1, D2, D3 was changed using a thermobalance.\\
\begin{table}
\caption{\label{table1}Overview of all YBCO samples with $T_{c}$ being their critical temperature and $T_{t}$ the temperature of the observed transition (n.d.: transition non-detectable).}
\begin{ruledtabular}
\begin{tabular}{lcclcc}
\multicolumn{3}{c}{\textit{polycrystalline samples}} & \multicolumn{3}{c}{\textit{single crystals}} \\
name & $T_{c}$ [K] & $T_{t}$ [K] &name & $T_{c}$ [K] & $T_{t}$ [K]\\
\cline{1-3} \cline{4-6}
C 6.19 & - & 340 $\pm$ 4 & D1 6.35 & - & n.d. \\
C 6.28 & - & 338 $\pm$ 2 & E 6.5 & 45 $\pm$ 1 & 335 $\pm$ 2 \\
C 6.34 & - & 338 $\pm$ 1 & F 6.6 & 55 $\pm$ 1 & n.d. \\
C 6.43 & 31 $\pm$ 1 & 337 $\pm$ 1 & D2 6.6 & 55 $\pm$ 1 & 334 $\pm$ 2 \\
C 6.52 & 55 $\pm$ 1 & 338 $\pm$ 3 & D3 6.7 & 70 $\pm$ 1 & n.d. \\
B 6.53 & 58 $\pm$ 1 & n.d. & D3 7.0 & 86 $\pm$ 1& n.d. \\
B 6.56 & 60 $\pm$ 1 & n.d. & & & \\
A 6.60 & 61 & n.d. & & &\\
A 6.68 & 62 $\pm$ 1 & n.d. & & &\\
A 6.73 & 68 $\pm$ 1 & n.d. & & &\\
C 6.79 & 80 $\pm$ 1 & 338 $\pm$ 1 & & &\\
C 6.90 & 91 $\pm$ 1 & 337 $\pm$ 1 & & &\\
C 7.0 & 91 $\pm$ 1 & 338.6 $\pm$ 1 & & &\\
\end{tabular}
\end{ruledtabular}
\end{table}
In the following, the magnetization M will be given in emu/mol and is normalized to the amount of YBCO in mol. 
Every measurement was started by a zero field-cooling (ZFC) of the sample. This was followed by a measurement during warming (FW) and cooling (FC) the sample under the same magnetic field. It should be noted that the magnetic irreversibility of the superconducting wire used for the fabrication of the magnetometers solenoid produces a small remanent magnetic field in the range of $\pm$10 Oe. Prior to the measurements at low field, this remanent field was determined by calibration of the magnetometer with a Pd sample and the field corrected for the remanent field will be referred to as  $H_{real}$.

\section{\label{results}Results}
\begin{figure}
\includegraphics[width=0.43\textwidth]{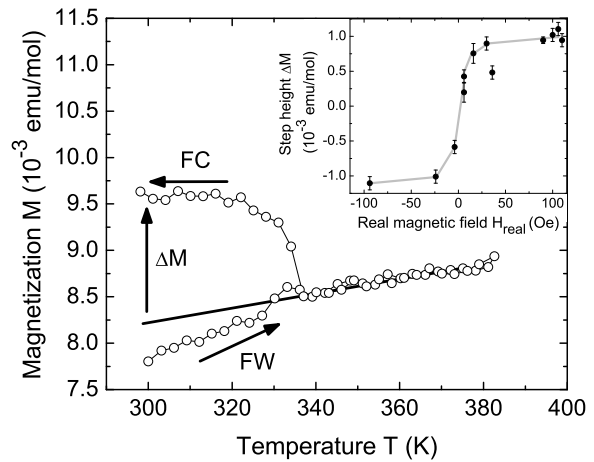}
\caption{\label{fig:figure1}Field warmed (FW) and field cooled (FC) magnetization of the C 6.43 polycrystalline YBCO sample as a function of temperature measured under a magnetic field of $H_{real}=$ 34 Oe after a zero field cooling (ZFC) procedure.  Inset: Step height $\Delta M$ of the observed transition for the same sample as a function of the real applied magnetic field $H_{real}$ obtained through the same ZFC-FW-FC procedure.}
\end{figure}

Figure \ref{fig:figure1} depicts the magnetization of a $x=6.43$ polycrystalline YBCO sample (C 6.43, see Tab. \ref{table1}) measured under a magnetic field of 34 Oe as a function of temperature in the range from 300 to 380 K. The FW magnetization $M_{FW}$ after a zero field-cooling increases linearly with increasing temperature up to $T\approx330$ K followed by a second slightly shallower linear increase for $T \geq 340$ K. The FC magnetization $M_{FC}$ exhibits a behavior similar to $M_{FW}$ for $T \geq 340$ K, but then shows a visible increase of the magnetization at a temperature of $T_{t}=(337 \pm 1)$ K. In order to investigate this transition quantitatively, we extrapolated the $M(T)$ curve above the transition to 300 K and defined the difference of the FC magnetization at 300 K with respect to this extrapolation as the step height $\Delta M$ of the transition. The step height $\Delta M$ of the transition shown in Fig. \ref{fig:figure1} amounts to (1.3 $\pm$ 0.1) x 10$^{-3}$ emu/mol.\\
The inset of Fig. \ref{fig:figure1} depicts $\Delta M$ as a function of the real applied field $H_{real}$ for the same sample, i.e., the remanent magnetic field in the magnetometers solenoid has been taken into account for these measurements. At first, the step height $\Delta M$ increases steeply with increasing magnetic field up to $H_{real}=\pm 30$ Oe and then saturates at $\Delta M\approx\pm 1$ emu/mol. Since the absolute value of the step height moves within the range of $10^{-6}$ emu, the transition could not be observed in high field measurements. The temperature of the transition $T_{t}$ showed no dependence on the magnetic field up to 0.1 T and is given in Tab. \ref{table1} for each sample.\\
\begin{figure}
\includegraphics[width=0.43\textwidth]{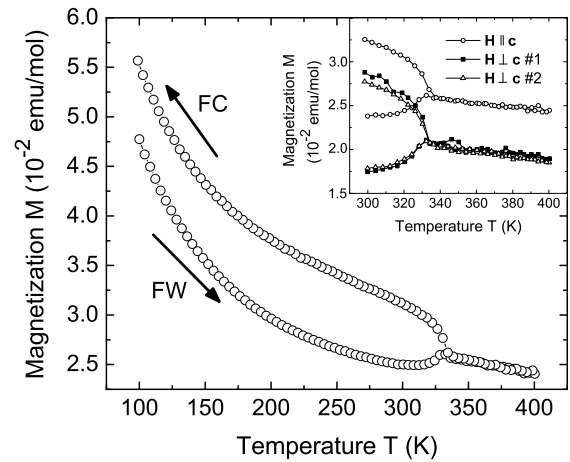}
\caption{\label{fig:figure2}Field warming (FW) and field cooling (FC) magnetization  of the E 6.5 single crystalline YBCO sample as a function of temperature measured under a magnetic field of 50 Oe with \textbf{H}$\parallel$\textbf{c}. Inset: Comparison of the magnetization of the same sample measured under a magnetic field of 50 Oe for \textbf{H}$\parallel$\textbf{c}, \textbf{H}$\perp$\textbf{c} \#1 and \textbf{H}$\perp$\textbf{c} \#2. The orientations \textbf{H}$\perp$\textbf{c} \#1 and \textbf{H}$\perp$\textbf{c} \#2 correspond to arbitrary perpendicular directions within the \textit{ab}-plane and not to the \textbf{a} and \textbf{b} axis of the sample. The initial field cooling was performed here under $\approx -10$ Oe resulting in a negative step height in $M_{FW}(T)$.}
\end{figure}
Figure \ref{fig:figure2} shows the magnetization of a $x=6.5$ YBCO single crystal (E 6.5) as a function of temperature measured under a field of 50 Oe with \textbf{H} parallel to the crystals \textbf{c} axis. The  $M(T)$ behavior is dominated by a strong Curie-Weiss contribution of paramagnetic secondary phases such as the green phase Y$_{2}$BaCuO$_{5}$ \cite{Parkin:1988}, BaCuO$_{2}$ \cite{Parkin:1988} and a Curie contribution from free Cu atoms in the Cu-chains of YBCO \cite{Leridon:2009}. In addition, $M_{FC}$ shows a visible transition at a temperature $T_{t}=(332 \pm 2)$ K. 
The inset of Fig. \ref{fig:figure2} depicts the magnetization of the $x=6.5$ single crystal (E 6.5) measured under a field of 50 Oe for three different orthogonal orientations of the crystal with respect to the magnetic field, clearly revealing the YBCO anisotropy of $M(T)$ between the directions  \textbf{H}$\parallel$\textbf{c} and  \textbf{H}$\perp$\textbf{c}. However, for all three orientations $M_{FC}$ shows a visible transition at $T_{t}=(332 \pm 2)$ K  with step heights of $\Delta M=(0.6 \pm 0.1)$ x $10^{-2}$ emu/mol for \textbf{H}$\parallel$\textbf{c} and \textbf{H}$\perp$\textbf{c} \#2 and $(0.7 \pm 0.1)$ x $10^{-2}$ emu/mol for \textbf{H}$\perp$\textbf{c} \#1, i.e., there seems to be no visible dependence of the step height on the orientation of the crystal with respect to the magnetic field.\\
We measured three different batches of polycrystalline YBCO samples and four different batches of single crystals with different oxygen contents $x$. The transition was only observed for one of the three batches of polycrystalline samples and two of the four batches of single crystals (see Tab. \ref{table1}), therefore ruling out the origin for the transition as intrinsic to YBCO. Neither the existence, nor the temperature of the transition did depend on the oxygen content between $x=0.19$ to $x=1.0$.  (An extensive study about the effects of oxygen disorder on the superconducting and pseudogap properties of YBCO was performed and is described elsewhere \cite{Biscaras:2012}. The present transition temperature was shown to be independent of these effects to within our measurement accuracy.) The average temperature $T_{t}$ of all detected transitions amounts to $(336 \pm 3)$ K.

\section{\label{disc}Discussion}
\begin{figure}
\includegraphics[width=0.43\textwidth]{{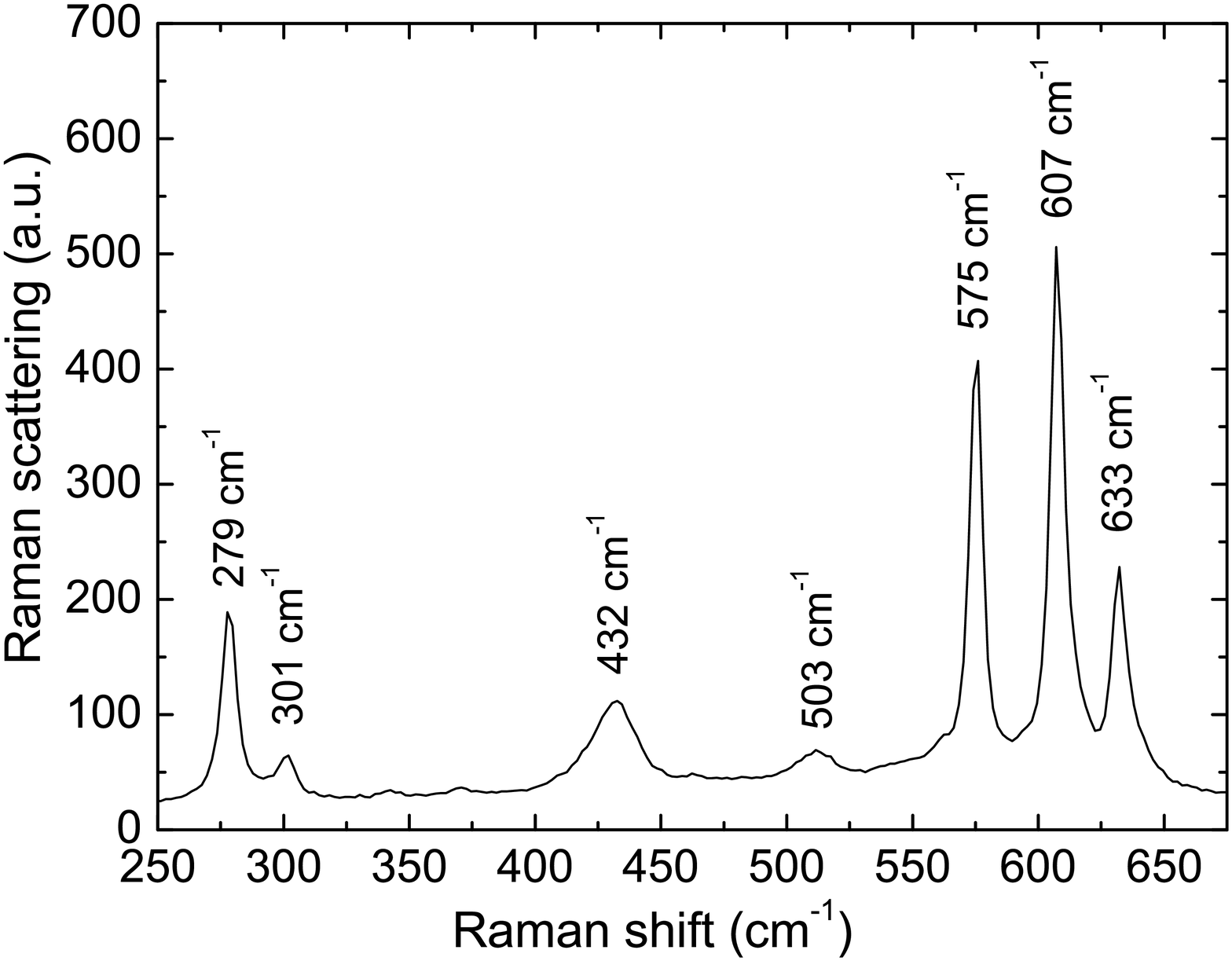}}
\caption{\label{raman}  Raman spectrum of the E 6.5 single crystalline YBCO sample.  This spectrum was obtained using a micro-Raman set-up and a T64000 triple monochromator. The 514.52 nm line of an Ar-Kr Laser was used.  The peaks at 279, 503, 575 and 633 cm$^{-1}$ were attributed to BaCu$_3$O$_4$ (see B. G\"uttler et al., Physica C 324, 123 (1999)). The 
peak at 607 cm$^{-1}$ was attributed to Y$_{2}$BaCuO$_{5}$ (see H. Rosen et al., Phys. Rev. B 36, 726 (1987))
whereas the peaks at 301 and 432 cm$^{-1}$ could not be identified. Similar spectra were obtained only at selected locations on the different samples.}
\end{figure}

\begin{figure}
\includegraphics[width=0.45\textwidth]{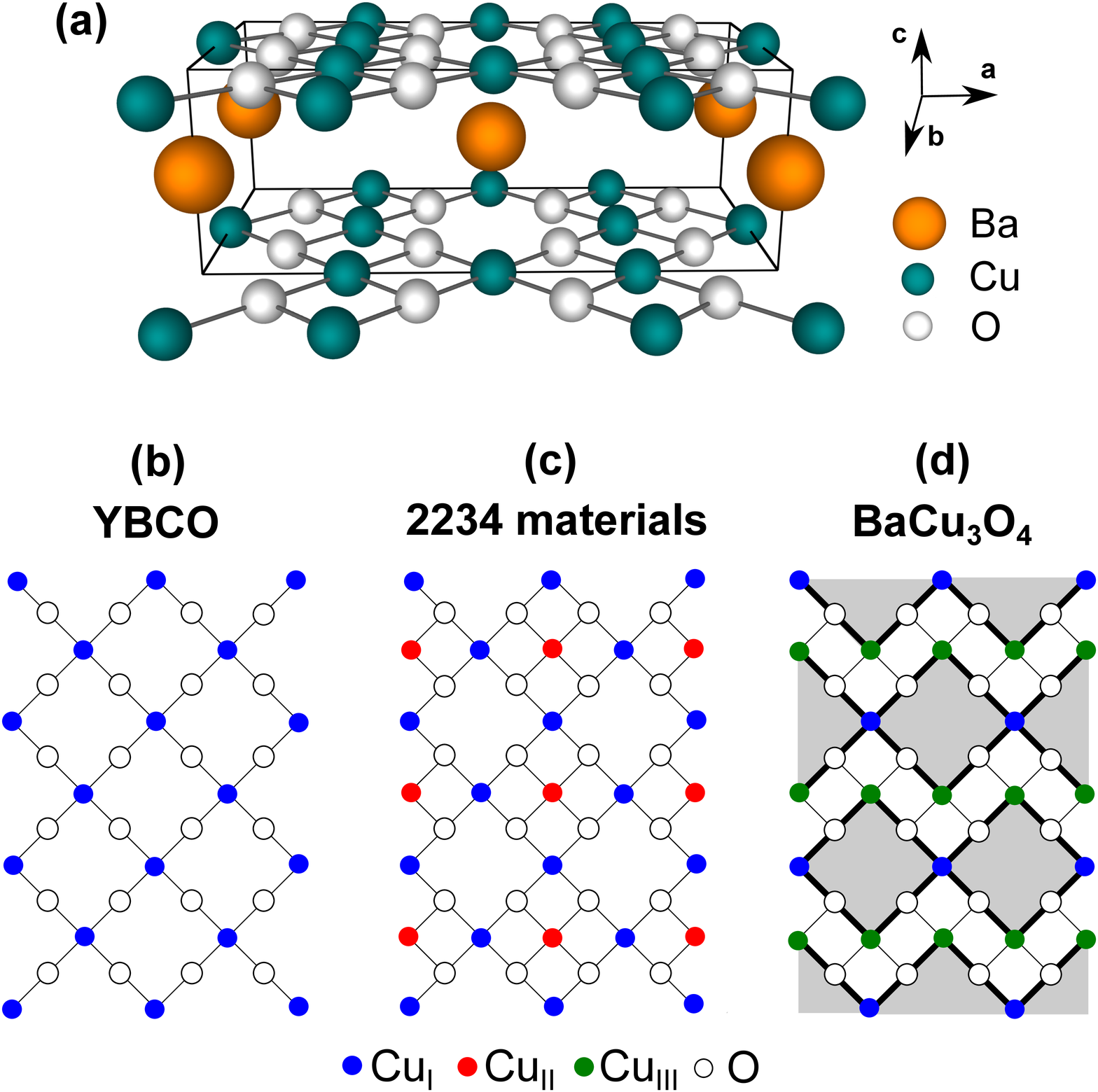}
\caption{\label{fig:figure3}\textbf{(a)} Crystal structure of BaCu$_3$O$_4$ \cite{Yakhou:1996}. \textbf{(b)}  CuO$_{2}$ planes in YBa$_2$Cu$_3$O$_{x}$. \textbf{(c)} Cu$_{3}$O$_{4}$ planes in Ba$_2$Cl$_2$Cu$_3$O$_4$ or Sr$_2$Cl$_2$Cu$_3$O$_4$. \textbf{(d)} Cu$_{3}$O$_{4}$ planes in BaCu$_{3}$O$_{4}$.} 
\end{figure}

Impurity phases present in YBCO  have been extensively studied. However, the average temperature of the observed transition of $(336 \pm 3)$ K does not match with any of the N\'eel or Curie temperatures of common impurity phases such as Y$_{2}$BaCuO$_{5}$ ($T_{N}=30$ K \cite{Parkin:1988}), Y$_{2}$Cu$_{2}$O$_{5}$ ($T_{N}=13$ K \cite{Troc:1987}), BaCuO$_{2}$ ($T_{N}=15$ K \cite{Wang:1995}) or CuO ($T_{N1,N2}=213, 230$ K \cite{Yang:1989}). This suggests that the transition is either due to an unknown impurity phase or an impurity phase with unknown magnetic properties.\\
As a matter of facts, we detected the presence of the alkaline earth oxocuprate BaCu$_{3}$O$_{4}$ in  samples showing the magnetic transition at $(336 \pm 3)$ K using Raman spectroscopy \cite{Guettler-1999,Rosen-1987}. The micro-Raman spectrum is displayed in Figure \ref{raman}.
BaCu$_{3}$O$_{4}$ was identified for the first time as a secondary phase in YBCO samples by Bertinotti \textit{et al.} using X-ray diffraction (XRD) \cite{Bertinotti:1989} and confirmed later \cite{Yakhou:1996}. Since BaCu$_{3}$O$_{4}$ is metastable and only grows epitaxially on YBCO or other perovskite structures, there are only few studies (among them Raman spectroscopy \cite{Guettler-1999,Rosen-1987}) of this material and its magnetic properties are unknown. However, as pointed out in Ref. \cite{Bertinotti:1989}, the structure of BaCu$_{3}$O$_{4}$ is close to the structure of the stable materials Ba$_2$Cl$_2$Cu$_3$O$_4$ and Sr$_2$Cl$_2$Cu$_3$O$_4$ (2234 materials) which show a  magnetic transition at 330 K \cite{Ito:1997} and 380 K \cite{Chou:1997}, respectively. BaCu$_3$O$_4$ consists of Ba and Cu$_3$O$_4$ planes (see Fig. \ref{fig:figure3} (a)) whereas the latters are composed of Ba (respectively Sr) planes, Cl planes and Cu$_3$O$_4$ planes. Fig. \ref{fig:figure3} depicts the CuO$_2$ planes of YBCO (b) and the Cu$_3$O$_4$ planes of 2234 materials (c) and BaCu$_3$O$_4$ (d). The CuO$_2$ layers of YBCO (see Fig. \ref{fig:figure3} (b)) are composed of a square lattice with Cu$^{2+}$ ions (d$^9$, S=1/2) on the corners (Cu$_{I}$ sites) and O ions on the edges. Nearest neighbor (NN) Cu$_I$ ions are coupled via superexchange interaction through the oxygen atom between them resulting in an antiferromagnetic (AF) order below 420 K for YBCO.\\
Regarding the similarities in the structure of the Cu$_3$O$_4$ planes, the magnetic properties of BaCu$_3$O$_4$ and 2234 materials are expected to bear some resemblances. The Cu$_3$O$_4$ planes of 2234-materials (see Fig. \ref{fig:figure3} (c)) consist of the CuO$_2$ structure shown in Fig. \ref{fig:figure3} (b) with additional Cu ions at the center of every second square (Cu$_{II}$ sites). The Cu$_I$-O-Cu$_I$ coupling (superexchange energy $J_{0}\approx 130$ meV) is similar to that described for the CuO$_2$ planes of YBCO and results in an AF order below $T_N\approx 330$ and 380 K for Ba$_2$Cl$_2$Cu$_3$O$_4$ \cite{Ito:1997} and Sr$_2$Cl$_2$Cu$_3$O$_4$ \cite{Chou:1997}, respectively. The Cu$_{II}$-Cu$_{II}$ coupling is much weaker ($J_{1}\approx 10$ meV) and gives rise to an AF order below $T_N\approx 30$ and 40 K for Ba$_2$Cl$_2$Cu$_3$O$_4$ \cite{Ito:1997} and Sr$_2$Cl$_2$Cu$_3$O$_4$ \cite{Chou:1997}, respectively, while no magnetic anomaly is observed in this range of temperature for our samples. Since the interaction of a Cu$_{II}$ ion with the four surrounding Cu$_I$ ions is frustrated, the Cu$_I$ and Cu$_{II}$ sublattices should be decoupled, except for quantum fluctuations \cite{Shender:1982}. The Cu$_3$O$_4$ planes of BaCu$_3$O$_4$ consist of Cu ions on two different sites, namely Cu$_I$ and Cu$_{III}$, and the lattice can be described with "diamond chains" (marked in gray in Fig. \ref{fig:figure3} (d)) consisting of squares attached by their Cu$_I$ corner ions. Cu$_I$ are strongly coupled to their four next NN Cu$_{III}$ via 180$^{\circ}$ Cu$_I$-O-Cu$_{III}$ ($J_0$) superexchange interaction and only weakly coupled to their two NN Cu$_{III}$.\\
Since the temperature of the observed transition $T_t=(336\pm3)$ K is close to $T_{N}^{I}$ of Ba$_2$Cu$_3$O$_4$Cl$_2$, we attribute $T_t$ to an AF order resulting from the 180$^{\circ}$ Cu$_I$-O-Cu$_{III}$ coupling. Unlike the Cu$_I$ and Cu$_{II}$ sublattices in 2234 materials, the Cu$_I$ and Cu$_{III}$ sublattices in BaCu$_3$O$_4$ are strongly coupled, resulting in a unique AF order temperature for all Cu ions. Another important difference with respect to the 2234 materials is that in BaCu$_3$O$_4$ the order is strictly \textit{one-dimensional} and not two-dimensional. It is noteworthy that the net magnetization of a \textit{single} diamond chain is not zero, but one Bohr magneton per three Cu ions since the cell pattern consists of three Cu ions, one bearing an opposite magnetization with respect to the two others.\\
\begin{figure}
\includegraphics[width=0.45\textwidth]{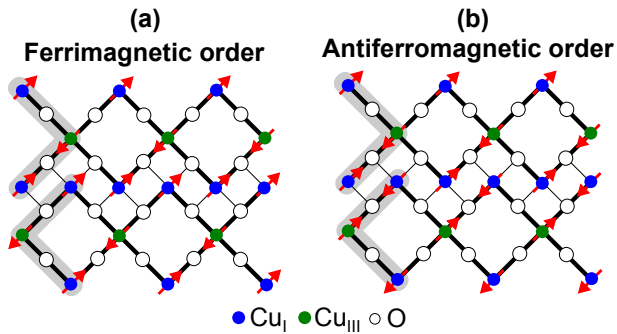}
\caption{\label{fig:figure4}Illustration of the two possible magnetic orientations of two neighbouring diamond chains. \textbf{(a)} Nearest neighbour chains are ferrimagnetically ordered and \textbf{(b)} nearest neighbour chains are  antiferromagnetically ordered.}
\end{figure}
There remains the question of the relative ordering of two adjacent and weakly coupled diamond chains. As weak as this interchain coupling may be, it favors the existence of one-dimensional order, which is a key issue inasmuch as such Ising-type order is forbidden when only a pure superexchange mechanism is present (another favoring mechanism may also be the spin-orbit coupling). The first hypothesis for the relative ordering of two adjacent chains is illustrated in Fig. \ref{fig:figure4} (a) and assumes that two NN diamond chains have parallel magnetizations. This would result in a \textit{ferrimagnetic} order and the net magnetization of one Cu$_3$O$_4$ layer would be 1/3$\mu_B$ per Cu. In the case where all layers are similarly oriented, we can estimate the molar fraction of BaCu$_3$O$_4$ in our samples to about 2$\times$10$^{-6}$, which is much too small to be detected by Raman or XRD spectroscopy and therefore not realistic.\\
A second hypothesis is that the chains have parallel magnetizations within one given layer but the layers are AF ordered. This could in principle give rise to metamagnetism with a "spin-flop" transition at sufficiently high magnetic fields which was not observed up to 5T.\\
The third hypothesis is that the magnetic order of two adjacent chains is antiparallel (see Fig. \ref{fig:figure4} (b)), so that the net in-plane magnetization is zero at zero field and the resulting order within the planes is \textit{antiferromagnetic}. In this case the observed magnetization can be due to the Dzyaloshinskii-Moriya (DM) interaction, which is allowed by symmetry (since the oxygen atom at the center of the Cu$_I$-O-Cu$_{III}$ bond does \textit{not} bear a center of inversion symmetry) and produces a net out-of-plane moment, able to rotate around the Cu-O bond axis under a magnetic field. This hypothesis of weak ferromagnetism is in fact the only one to be consistent with our observation that the step height $\Delta M$ is of the same amplitude for all orientations of the crystal with respect to the magnetic field (see Fig. \ref{fig:figure2}) (note that at zero field the DM contributions of two neighboring chains cancel out).  If the third hypothesis is valid, then making the rough assumption that the magnetic step heights $\Delta M$ in 2234 materials and in BaCu$_3$O$_4$ are the same, the molar fraction of BaCu$_3$O$_4$ in our samples would be about 2$\times$10$^{-3}$, therefore compatible with Raman observation.\\
Through neutron scattering measurements on underdoped YBCO, Sidis \textit{et al.} \cite{Sidis:2001} reported coexistence of magnetism and superconductivity at the doping level $x=6.5$. Part of the sample used for this study is our sample E 6.5 in which we detected BaCu$_3$O$_4$ through Raman spectrometry (see Figure \ref{raman}). The above inferred concentration could also be compatible with the observation of a very small moment through neutron scattering, but too small to be detectable through muon-spin-resonance ($\mu$SR) \cite{Sidis:2001}.  Since BaCu$_3$O$_4$ grows epitaxially on YBCO, this also explains the commensurability of the reported magnetic order. This newly discovered magnetic transition in BaCu$_3$O$_4$ might then account for some cases of apparent coexistence of magnetism and superconductivity in underdoped YBCO.

From the theoretical point of view, according to the pioneering work of Takano and coworkers \cite{Takano:1996} on the phase diagram of diamond chains, BaCu$_3$O$_4$ appears indeed to belong to the ferrimagnetic case, however chain interactions deserve further study.

\section{\label{conclusion}Conclusion}
We reported on the magnetic properties of BaCu$_3$O$_4$, a common parasitic phase in YBa$_2$Cu$_3$O$_x$ samples. The magnetic order that sets in at 336 K is most probably due to diamond chains of antiferromagnetically coupled Cu 1/2 spins. The measured magnetization may be attributed to either ferrimagnetic ordering or (most probably) anti-ferromagnetic ordering of adjacent chains in the presence of the Dzyaloshinsky-Moriya interaction yielding weak ferromagnetism behavior. These diamond chains represent a new remarkable system of weakly coupled one-dimensional magnetic objects where AF ordering of the Cu$^{2+}$ ions in each diamond chain leads to a net magnetic moment of each chain.

\begin{acknowledgments}
\textbf{Acknowledgments:}\
This work was supported through a SESAME grant from Region Ile-de-France. We gratefully thank P. Bourges, Y. Gallais,  A. Sacuto, X. Chaud and F. Mila for stimulating discussions or experimental collaboration. 
\end{acknowledgments}

\end{document}